\documentclass[twocolumn,superscriptaddress,showpacs,aps,floatfix]{revtex4-1}
\usepackage{amsmath,amsfonts,amssymb,epsfig,dcolumn,bm,dsfont,graphics,latexsym,color,graphicx}
\let\oldAA\AA

\renewcommand{\AA}{\text{\normalfont\oldAA}}
\newcommand{\ket}[1]{| {#1} \rangle} % for Dirac bras
\begin{document}
\preprint{AIP/123-QED}
\title{Proximity-induced spin-orbit coupling in phosphorene on a WSe$_2$ monolayer}
\author{Marko Milivojevi\'c}
\affiliation{Institute of Informatics, Slovak Academy of Sciences, 84507 Bratislava, Slovakia}
\affiliation{Institute for Theoretical Physics, University of Regensburg, 93053 Regensburg,
Germany}
\affiliation {Faculty of Physics, University of Belgrade, 11001 Belgrade, Serbia}
\author{Martin Gmitra}
\affiliation{Institute of Physics, Pavol Jozef \v{S}af\'{a}rik University in Ko\v{s}ice, 04001 Ko\v{s}ice, Slovakia}
\affiliation{Institute of Experimental Physics, Slovak Academy of Sciences, 04001 Ko\v{s}ice, Slovakia}
\author{Marcin Kurpas}
\affiliation{Institute of Physics, University of Silesia in Katowice, 41‑500 Chorz\'ow, Poland}
\author{Ivan \v Stich}
\affiliation{Institute of Informatics, Slovak Academy of Sciences, 84507 Bratislava, Slovakia}
\affiliation{Department of Natural Sciences, University of Saints Cyril and Methodius, 917 01 Trnava, Slovakia}
\author{Jaroslav Fabian}
\affiliation{Institute for Theoretical Physics, University of Regensburg, 93053 Regensburg, Germany}
%\date{\today}
%%%%%%%%%%%%%%%%%%%%%%%%%%%%%%%%%%%%%
\begin{abstract}
We investigate, using first-principles methods and effective-model simulations, the spin-orbit coupling proximity effects in a bilayer heterostructure comprising phosphorene and WSe$_2$ monolayers.  We specifically analyze holes in phosphorene around the $\Gamma$ point, 
at which we find a significant increase of the spin-orbit coupling that can be attributed to the strong hybridization of phosphorene with the WSe$_2$ bands. We also propose an effective spin-orbit model based on the ${\bf C}_{1{\rm v}}$ symmetry of the studied heterostructure. 
The corresponding spin-orbit field can be divided into two parts: the in-plane field, present due to the broken nonsymmorphic horizontal glide mirror plane symmetry, and the dominant out-of-plane field triggered by breaking the out-of-plane rotational symmetry of the phosphorene monolayer. Furthermore, we also demonstrate that a heterostructure with 60$^\circ$ twist angle exhibits an opposite out-of-plane spin-orbit field, indicating that the coupling can effectively be tuned by twisting. The studied phosphorene/WSe$_2$ bilayer is a prototypical low common-symmetry heterostructure in which the proximity effect can be used to engineer the spin texture of the desired material.
\end{abstract}
%%%%%%%%%%%%%%%%%%%%%%%%%%%%%%%%%%%%%
\maketitle
\section{Introduction}

Phosphorene~\cite{LNZ+14,LNH+14,LWL+14,ZYX+14,CVP+14,QKH+14,ZZS+15,ZZC+17} is a two-dimensional (2D) material whose sizable direct semiconducting gap and high carrier mobility make it a promising alternative to gapless graphene in the field of electronics. However, weak spin-orbit coupling~\cite{LA14,KGF16,JKG+19,KJG+19} and zero magnetism in phosphorene, limit its use in spintronics applications. Also, phosphorene has space-inversion symmetry and thus exhibits no 
spin-orbit fields. The simplest way to induce such fields is via the Rashba effect~\cite{PKS15,FR19}, i.e., by applying an electric field in the direction perpendicular to the monolayer plane. This approach is not very effective in phosphorene as the Rashba field ultimately depends on the atomic number~\cite{SPS14}. It is therefore desired to find alternative ways of inducing sizable spin-orbit fields in phosphorene. 

Van der Waals heterostructures offer a rich playground for modifying electronic, spin, optical, and magnetic properties of the target materials~\cite{GG13,HMM+20,SFK+21,ZZN+22,HZL+22,SBB+22,ZRM+23}. In the context of proximity-induced spin-orbit effects in weak SOC materials~\cite{GF15,GF17,SMK+22}, transition-metal dichalcogenide (TMDC) monolayers (MLs)~\cite{MLH+10,KH12,CRG+13} are the obvious material of choice due to the strong spin-orbit coupling of their valence bands~\cite{ZCS11,KZD+13,SYZ+13,KGR13,ABX+14,KBG+15}.

The common three-fold symmetry of graphene and TMDC materials has enabled a simple effective description of the proximity-induced interaction between the  MLs~\cite{LK19,VGR21,NZG+21,LSK+22,DRK+19,PDR+22}. 
Such a common symmetry is not present in phosphorene/TMDC heterostructures, in which the rotation-symmetry-broken environment can trigger different spin-orbit coupling terms and, as a consequence, induce new types of spin textures in the desired materials~\cite{JOS+21}. 

The goal of the present study is to obtain both a quantitative and qualitative understanding of such heterostructures. In particular, we study
a heterostructure comprising phosphorene (P) and monolayer WSe$_2$ employing \emph{ab-initio} methods and group theory. The giant spin splitting in the valence bands of the WSe$_2$ monolayer points to the potentially interesting hole spin physics of proximitized phosphorene. Indeed, we find sizable momentum-dependent spin-orbit fields at the $\Gamma$ point (both in-plane and out-of-plane) where the strong hybridization between the phosphorene and WSe$_2$ bands takes place. From symmetry arguments, we derived an effective spin-orbit Hamiltonian that ideally captures the spin physics predicted by the density-functional theory (DFT) calculations. Finally, we show that a 60$^\circ$ twisted heterostructure preserves the in-plane spin-orbit fields but flips the out-of-plane component, suggesting that twist angle can be an effective tool to tailor the proximity spin physics in such heterostructures. 

This paper is organized as follows. After the introductory section, in  Sec.~\ref{Heterostructure}, we analyze the geometry of the P/WSe$_2$ heterostructure and present the necessary computational details for the calculation of the band structure. In Sec.~\ref{bandstructure}, band structure analysis of such a heterostructure is presented. Furthermore, based on the ${\bf C}_{1{\rm v}}$ symmetry of the heterostructure, the effective model for the hole spins around the $\Gamma$ point is constructed and the fitting parameters that match the DFT data with the model are given. We also analyze the effect of twist on the proximity effect, by assuming the relative twist angle of $60^{\rm o}$ between the phosphorene and WSe$_2$ monolayer. Finally, in Sec.~\ref{Conclusions}, we present our conclusions and provide further outlooks of the presented study.

%%%%%%%%%%%%%%%%%%%%%%%%%%%%%%%%%%%%%%%%%%%%%%%%%%%%%%%%%%%%%%%%%%%%%%%%%%%%%%%
\section{Computational and atomic structure details}\label{Heterostructure}

\begin{figure}[t]
\centering
\includegraphics[width=0.48\textwidth]{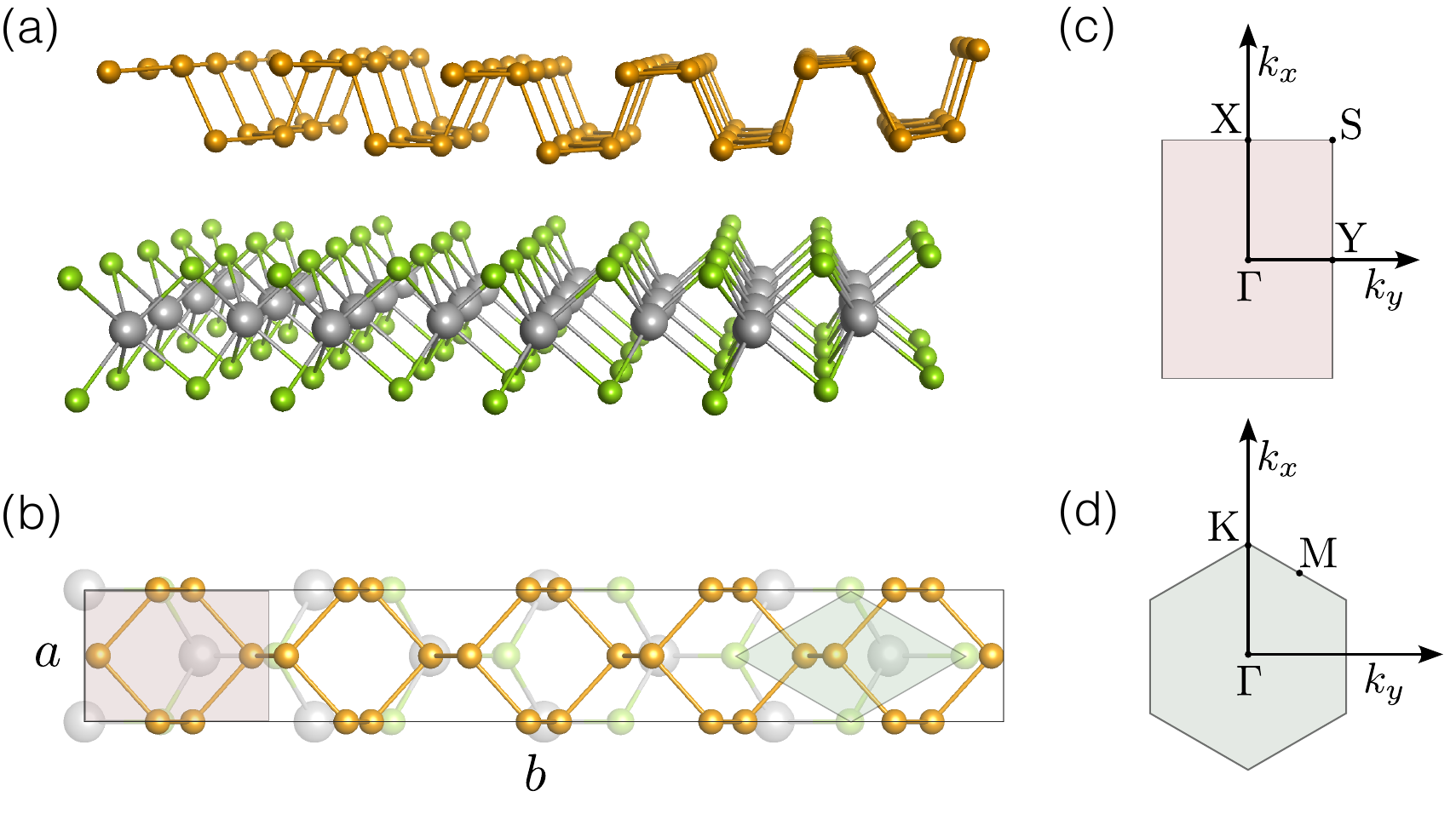}
\caption{Atomic structural model of studied P/WSe$_2$ heterostructure. (a)~Side perspective view and (b)~top view with primitive unit cells of phosphorene and WSe$_2$ are shaded in gray. In~(c) and~(d) the Brillouin zones with high symmetry points of phosphorene and WSe$_2$ monolayer is also given. We identify the $x$/$y$ direction of the heterostructure with the zigzag/armchair direction of the phosphorene monolayer. }\label{FigHet}
\end{figure}

For lattice parameters of phosphorene ML, we consider $a=3.2986\AA$ and $b=4.6201\AA$~\cite{JKG+19} (lattice vectors correspond to ${\bm a}=a {\bf e}_x$, ${\bf b}=b {\bf e}_y$), while the lattice parameter of WSe$_2$ ML is equal to $a_{{\rm W}}=3.286\AA$~\cite{WJ69} (lattice vectors are ${\bm a}_1=a_{\rm W} {\bf e}_x$,  ${\bm a}_2=a_{\rm W}(-{\bf e}_x+\sqrt{3}{\bf e}_y$)/2). 
The commensurate heterostructure was constructed using the CellMatch code~\cite{L15}, containing 20 P atoms and 8 WSe$_2$ chemical units.
While the phosphorene layer remained unstrained, the WSe$_2$ is strained by 0.51\%. In FIG.~\ref{FigHet}, we present a side~(a) and top~(b) view of the atomic structure model of the P/WSe$_2$ heterostructure, alongside the Brillouin zone with high symmetry points of phosphorene~(c) and WSe$_2$~(d) ML. The studied heterostructure has the vertical mirror plane symmetry that coincides with the $yz$ plane, where the zigzag (armchair) direction of phosphorene corresponds to the $x$ $(y)$ direction of the heterostructure. 
%%%%%%%%%%%%%%%%%%%%%%%%%%%%%%%%%%%%%%%%
%%%%%%%%%%%%%%%%%%%%%%%%%%%%%%%%%%%%%%%%%%%%%%%%%%%%%%%

We perform DFT electronic structure calculations of the P/WSe$_2$ heterostructure by means of the plane wave QUANTUM ESPRESSO package~\cite{QE1,QE2}, assuming a vacuum of 20\,${\AA}$ in the $z$-direction. The Perdew–Burke–Ernzerhof exchange-correlation functional was utilized~\cite{PBE96}, for the norm-conserving method~\cite{HSC79}.
The positions of atoms were relaxed with the help of the quasi-Newton scheme and scalar-relativistic SG15 Optimized Norm-Conserving Vanderbilt (ONCV) pseudopotentials~\cite{H13,SG15,SGH+16}. The force and energy convergence thresholds for ionic minimization were set to $1\times10^{-4}$~Ry/bohr and $10^{-7}$ Ry/bohr, respectively, using the Monkhorst-Pack scheme with $56\times 8$ $k$-points mesh.
Small Methfessel-Paxton energy level smearing of 1mRy~\cite{MP89} was used along with the kinetic energy cut-offs for the wave function and charge density 80\,Ry and 320\,Ry, respectively. Also, the semiempirical Grimme's DFT-D2 van der Waals (vdW) corrections were included~\cite{G06,BCF+09}. For the relaxed structure, the average distance between the closest phosphorene and the selenium plane (in the $z$-direction) is equal to $3.31\AA$. In the case of noncolinear DFT calculations including spin-orbit coupling, fully relativistic SG15 ONCV pseudopotentials were used. Also, the dipole correction~\cite{B99} was applied to properly determine energy offset due to dipole electric field effects. The energy convergence  threshold was set to $10^{-8}$~Ry/bohr, using the same $k$-points mesh and  kinetic energy cutoffs for the wave function and charge density as in the relaxation procedure. 

Note that the illustration of the band structure unfolded to the Brillouin zone of both monolayers, is done using the DFT Vienna ab-initio simulation package VASP~6.2~\cite{KF96,KF99}, using the relaxed structure from QUANTUM ESPRESSO code as the input.
%%%%%%%%%%%%%%%%%%%%%%%%%%%%%%%%%%%%%%%%%%%%%%%%%%%%%%%%%
%%%%%%%%%%%%%%%%%%%%%%%%%%%%%%%%%%%%%%%%%%%%%%%%%%%%%%%%%
%%%%%%%%%%%%%%%%%%%%%%%%%%%%%%%%%%%%%%%%%%%%%%%%%%%%%%%%%
\section{Band structure analysis}\label{bandstructure}
\begin{figure}[t]
\centering
\includegraphics[width=0.40\textwidth]{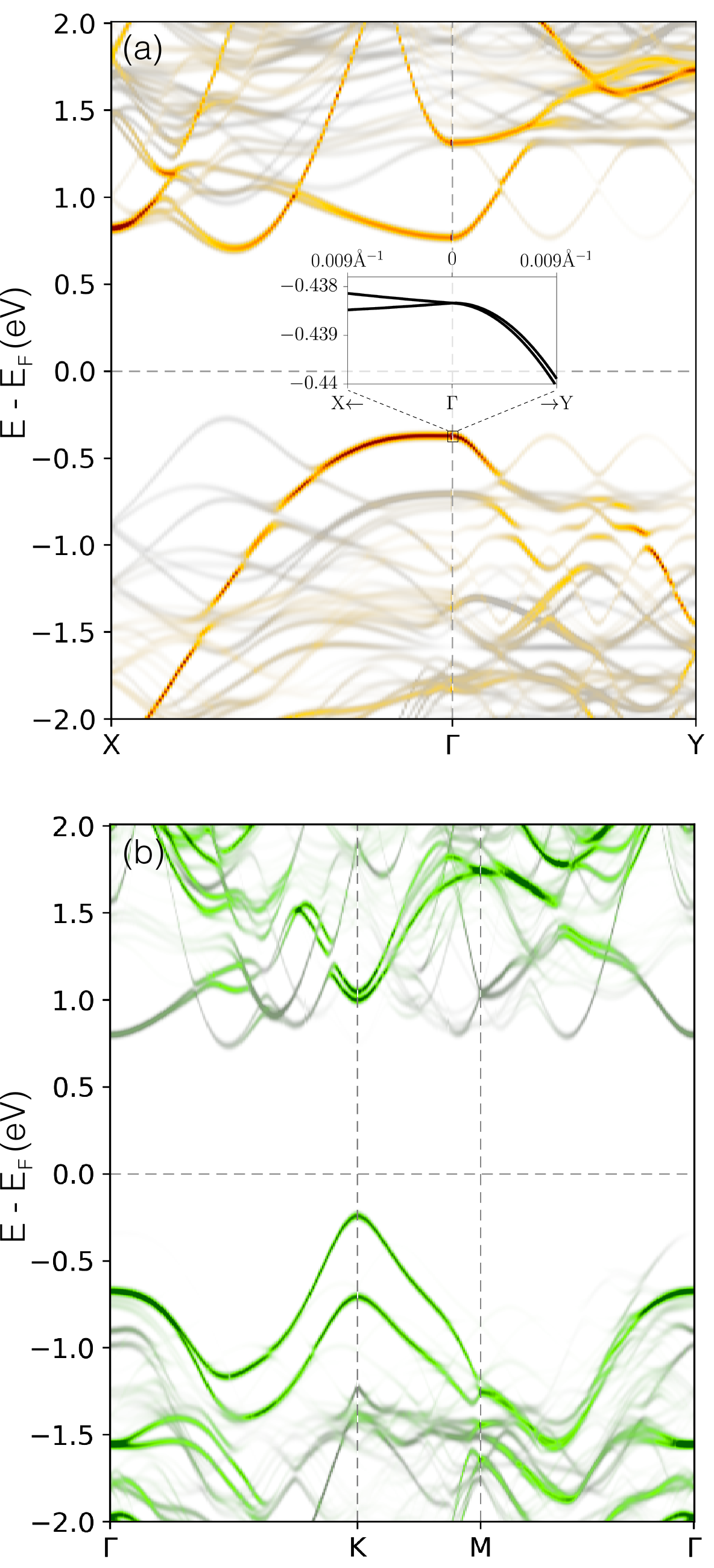}
\caption{Calculated band structure of a zero twist-angle commensurate P/WSe$_2$ heterostructure.
(a)~Unfolded band structure to the X$\Gamma$Y path of the phosphorene, inset shows detail of bands spin splitting;
(b)~band along the $\Gamma$KM$\Gamma$ path of the WSe$_2$ monolayer Brillouin zone. The bands with the dominant contribution of phosphorus~(a) and WSe$_2$~(b) atomic orbitals are marked with orange and green color, respectively.} \label{ORBITALcontribution}
\end{figure}

%%%%%%%%%%%%%%%%%%%%%%%%%%%%%%%%%%%
In FIG.~\ref{ORBITALcontribution} we present the band structure of the P/WSe$_2$ heterostructure unfolded to the X$\Gamma$Y path~(a) of the phosphorene and $\Gamma$KM$\Gamma$ path~(b) of the WSe$_2$ Brillouin zone.
Due to the reduced symmetry of the heterostructure with respect to their constituents, all the 
bands are spin split, as expected from the double group analysis~\cite{NMV+18}.
In order to have a more apparent separation between the bands having different atomic origins, we mark the bands with dominant phosphorus~(a) and WSe$_2$~(b) atomic orbital characters with orange and green color, respectively.
First, we notice that an overall heterostructure is a semiconductor due to the semiconducting nature of both constituents. 
The small strain applied to the WSe$_2$ monolayer does not change its band structure significantly. The most important feature for the spin-orbit proximity study stems from the fact that the top valence band projected to the WSe$_2$ Brillouin zone has the same characteristics as in the monolayer limit; the giant spin-orbit coupling at the K point and along the $\Gamma$KM path is preserved~\cite{ZCS11}. On the other hand, it can be seen that within the phosphorene Brillouin zone, the valence band around the $\Gamma$ point is mainly composed of phosphorene atomic orbitals. This is consistent with the highly anisotropic energy dispersion relation in the armchair and zigzag direction observed, resembling the well-known asymmetry of the phosphorene effective mass in the vicinity of $k=0$ point~\cite{FY14}. 
Additionally, close to the $\Gamma$ point, we notice strong hybridization of phosphorene bands with bands having dominant WSe$_2$ character. Since the K point of WSe$_2$ is folded to the X$\Gamma$ line of the phosphorene Brillouin zone, it is to be expected that the proximity-induced spin-orbit coupling should be more pronounced along the X$\Gamma$ line than in the $\Gamma$Y direction. The DFT calculation confirms this conjecture.
As we will show below,
the obtained hole spin texture of the top valence band of phosphorene can be described using a simple 
symmetry-adapted spin-orbit Hamiltonian with anisotropic parameters for 
$\Gamma$X and $\Gamma$Y directions.
%%%%%%%%%%%%%%%%%%%%%%%%%%%%%%%%%%
%%%%%%%%%%%%%%%%%%%%%%%%%%%%%%%%%%

\subsection{Model Hamiltonian}\label{ModelHamiltonian}
To make a simple description of the hole physics in phosphorene within the P/WSe$_2$ 
heterostructure we derive a simple spin-orbit coupling model Hamiltonian based on the ${\bf C}_{1{\rm v}}$ symmetry of the heterostructure. Group symmetry  ${\bf C}_{1{\rm v}}=\{e,\sigma_{\rm v}\}$ has two elements; $e$ represents the identity element, while $\sigma_{\rm v}$ is the vertical mirror symmetry that coincides with the $yz$-plane. The presence of vertical mirror symmetry is a consequence of the zero twist angle between the MLs. This symmetry can be broken by twisting WSe$_2$ ML for an angle different than a multiple of $60^{\rm o}$, and would lead to the general spin-orbit coupling Hamiltonian of the form
$H_{\rm SO}^{\rm gen}=\sum_{i=x,y,z}(\alpha_ik_x+\beta_i k_y)\sigma_i$, with six unknown parameters $\alpha_i$ and $\beta_i$, generating a different type of spin texture than in the $0^{\rm o}/60^{\rm o}$ twist angle case.

Thus, the effective spin-orbit model close to the $\Gamma$ point in the $0^{\rm o}/60^{\rm o}$ twist angle case can be derived using the constraints posed by the presence of the vertical mirror plane symmetry as well as by the time-reversal symmetry.
Using the transformation rule of the momentum and spin operators, $(k_x,k_y)\xrightarrow{\sigma_{\rm v}} (-k_x,k_y)$ and $(\sigma_x,\sigma_y,\sigma_z)\xrightarrow{\sigma_{\rm v}} (\sigma_x,-\sigma_y,-\sigma_z)$, respectively, it turns out that the effective, linear in $k$, spin-orbit coupling Hamiltonian can be written as a sum of polynomials $k_x\sigma_y$, $k_y\sigma_x$, and $k_x\sigma_z$, that are invariant under the system's symmetry $\sigma_{\rm v}$
%%%%%%%%%%%%%%%%%%%%%%%%%%%%%%%%%%%%%%%%%
\begin{equation}\label{Heff}
    H_{\rm SO}^{\rm eff}=\lambda_{1} k_x \sigma_y+\lambda_{2} k_y \sigma_x+\lambda_3 k_x \sigma_z,
\end{equation}
%%%%%%%%%%%%%%%%%%%%%%%%%%%%%%%%%%%%%%%%
with the parameters $\lambda_{1}$, $\lambda_{2}$,  
and  $\lambda_{3}$ that need to be determined. The
presence of the $k_x \sigma_y$ and $k_y \sigma_x$ terms is a consequence of a broken nonsymmorphic horizontal glide mirror plane symmetry of the phosphorene monolayer, while the emergence of the $k_x\sigma_z$ spin-orbit fields triggered by breaking the out-of-plane rotational symmetry. In terms of the induced spin texture,  the spin-orbit Hamiltonian can be divided into two parts, the in-plane ($\lambda_1k_x \sigma_y+\lambda_2k_y \sigma_x$), and out-of-plane ($\lambda_3k_x \sigma_z$) spin-orbit fields.

\begin{figure*}[t]
\centering
\includegraphics[width=0.85\textwidth]{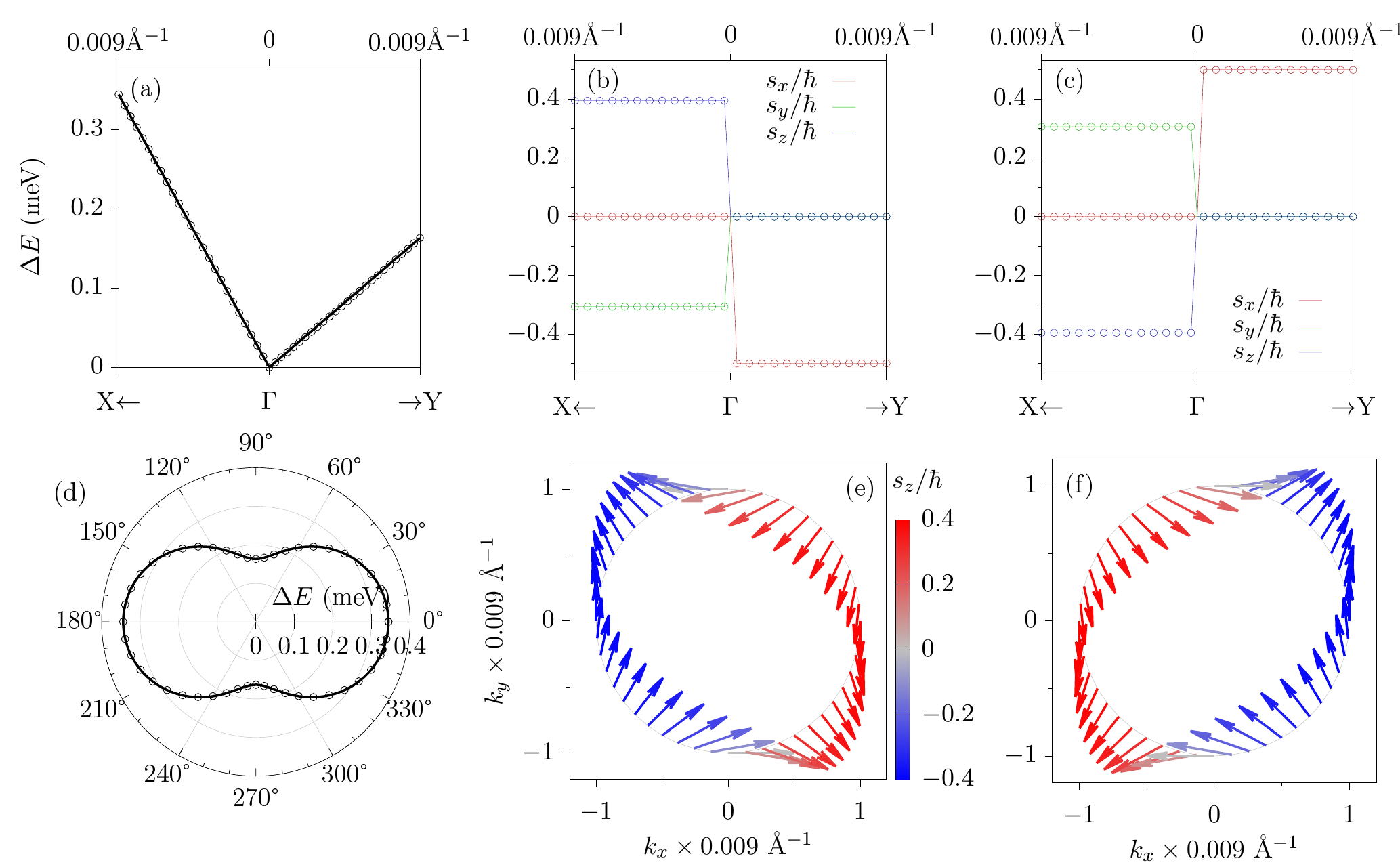}
\caption{Calculated electronic band spin splitting and spin expectation values for phosphorene top valence band in the P/WSe$_2$ heterostructure. 
(a)~Band spin splitting along the high symmetry lines in the first Brillouin zone;
(b)~spin expectation values for the lower band, and (c)~for the upper band along the high symmetry lines in the first Brillouin zone.
(d)~Angular dependence of the band spin splitting for the momenta around the $\Gamma$ point with radius $k=0.009~{\rm\AA}^{-1}$, 
(e)~spin expectation values for the lower and (f)~upper band spin split band. The color scale corresponds to the $z$-component of the spin.}\label{ODEG}
\end{figure*}

By diagonalizing the Hamiltonian~\eqref{Heff}, one can obtain the following formulas for the spin splitting and the spin expectation values of the Bloch states:
%%%%%%%%%%%%%%%%%%%%%%%%%%%%%%%%%%%%%%
\begin{eqnarray}
\Delta_{\rm so}^{\mp}&=&\mp \sqrt{k_x^2(\lambda_1^2+\lambda_3^2)+k_y^2\lambda_2^2},\nonumber\\
s_x^{\mp}&=&\mp\frac{k_y\lambda_2}{2\sqrt{k_x^2(\lambda_1^2+\lambda_3^2)+k_y^2\lambda_2^2}},\nonumber\\
s_y^{\mp}&=&\mp\frac{k_x\lambda_1}{2\sqrt{k_x^2(\lambda_1^2+\lambda_3^2)+k_y^2\lambda_2^2}},\nonumber\\
s_z^{\mp}&=&\mp\frac{k_x\lambda_3}{2\sqrt{k_x^2(\lambda_1^2+\lambda_3^2)+k_y^2\lambda_2^2}},
\end{eqnarray}
and use them to determine the spin-orbit coupling parameters by fitting the DFT data. The fitting parameters $\lambda_1=0.012$ eV\,\AA, $\lambda_2=0.009$ eV\,\AA, and $\lambda_3=-0.015$ eV\,\AA\, reproduce well the spin structure of the top valence band close to the $\Gamma$ point. This is illustrated in FIG.~\ref{ODEG}~(a)-(c) where we plot the spin-splitting energy $\Delta E=\Delta_{\rm so}^+-\Delta_{\rm so}^-$ and spin expectation values close to the $\Gamma$ point, along the
X$\Gamma$Y path. In FIG.~\ref{ODEG}(d)-(f), the angular dependence of spin splitting and spin expectation values is given by assuming the 
fixed $|\bf{k}|$ value (0.009 in the units of $1/\AA$, corresponding to the 0.94\% of the $\Gamma$X line and 1.32\% of the $\Gamma$Y line) and varying the angle $\varphi$ between the ${\bf k}$-point vector and the $x$-direction from 0 to 2$\pi$.
%%%%%%%%%%%%%%%%%%%%%%%%%%%%%%%%%%%%%%%%%%%%%%%%%%%%
%%%%%%%%%%%%%%%%%%%%%%%%%%%%%%%%%%%%%%%%%%%%%%%%%%%%

We first mention that the effective spin-orbit Hamiltonian model~\eqref{Heff} is valid up to 30\% of the $\Gamma$X path and 6\% of the $\Gamma$Y path, corresponding to the energies of 0.12~eV and 0.02~eV, respectively, from the valence band maximum.  The out-of-plane spin-orbit field is the inherent feature of the group-IV monochalcogenides~\cite{GC15,FLL+15,GCN15,PVL19,AMS+19} monolayers, representing the ferroelectrics with phosphorene-like atomic structure. However, in these systems, the spin is locked in the z-direction, due to symmetry, whereas in our case the more exotic spin texture is generated. Furthermore, one can compare the strengths of the spin-orbit coupling parameters in the $k_x$ and $k_y$ directions. In the $k_x$-direction, the effective strength of the spin-orbit field is equal to $\sqrt{\lambda_{1}^2+\lambda_{3}^2}=0.019$\,eV\AA~(comparable to the intrinsic spin-orbit coupling strength in ferroelectric SnS monolayer~\cite{AI19}), while in the $k_y$-direction, the strength is equal to $0.009$\,eV\AA, being roughly two times smaller than in the $k_x$ case.  

How can the proximity-enhanced spin-orbit coupling influence the electron spin dynamics in phosphorene?
We propose to explore spin relaxation, which is readily experimentally accessible. Indeed, in pristine phosphorene,
the spin relaxation was found from theory and experiment to be dominated by the Elliott-Yafet mechanism stemming from the intrinsic spin-orbit coupling \cite{KGF16,Avsar2017}. This competes with the Dyakonov-Perel mechanism, 
which is weaker due to the weak Rahsba spin-orbit coupling, although for sufficiently large out-of-plane electric fields or $z$-component of the crystal potential gradient $\nabla V(\bm{r})$, it can overtake the Elliott-Yafet effect. For monolayer phosphorene, this would happen for electric fields of $E \approx 5$\,Vnm$^{-1}$, corresponding to the effective strength of the spin-orbit field $\lambda_x \approx 1.08$\,meV\AA~in the $k_x$ direction and $\lambda_y \approx 3.34$\,meV\AA~in the $k_y$ direction \cite{KGF16}. The values of $\lambda_1$, $\lambda_2$, and $\lambda_3$ exceed those of $\lambda_x$ and $\lambda_y$. We thus
predict that the Dyakonov-Perel mechanism dominates the spin relaxation in proximitized phosphorene. 

From the comparison of spin-orbit coupling parameters, $\lambda$s', one sees that proximitized phosphorene due to WSe$_2$ has a pronounced anisotropy of the in-plane spin-orbit fields which is expected to yield marked spin relaxation anisotropy. Assuming the Fermi level is 2\,meV below the valence band maximum, the corresponding crystal momenta are $k_x = 0.015$\,\AA$^{-1}$ 
and $k_y = 0.0004$\,\AA$^{-1}$, which give spin-orbit fields $\Omega_x=\lambda_2 k_y = 3.6\,\mu$eV, $\Omega_y=\lambda_1 k_x = 0.18$\,meV and $\Omega_z = \lambda_3 k_x = 0.22$\,meV. It is clear, that $\Omega_x$ will have a minor effect on spin relaxation compared to $\Omega_y$ and $\Omega_z$. Neglecting $\Omega_x$, and assuming isotropic momentum lifetime $\tau_p$, the spin relaxation rates for the armchair (arm) and out-of-plane ($\perp$) directions the rates can be estimated as $\tau_{s,\rm arm}^{-1}\sim \tau_p  \lambda_3^2 \langle k_x^2\rangle $ and $\tau_{s,\perp}^{-1}\sim \tau_p  \lambda_1^2 \langle k_x^2\rangle$, respectively, where $\langle \rangle$ denotes the Fermi contour average~\cite{Zutic2004}. Electron spins polarized in the zigzag (zz) direction would relax approximately twice faster, with the rate $\tau_{s,\rm zz}^{-1}\sim \tau_p\langle k_x^2\rangle (\lambda_1^2+\lambda_3^2 )$. 

%%%%%%%%%%%%%%%%%%%%%%%%%%%%%%%%%%%%%%%%%%%%%%%%%%%%
%\begin{table}[t]
%\caption{Spin-orbit coupling parameters $\lambda_{\rm 1/2/3}$ obtained after fitting the model Hamiltonian~\eqref{Heff} to the DFT data (obtained using the Grimme-D2 vdW correction), assuming the P/WSe$_2$ heterostructure with zero twist angle.}\label{fits}
%\centering
%\setlength{\tabcolsep}{15pt}
%\renewcommand{\arraystretch}{1.2}
%\begin{tabular}{|c|c|c|}
%\hline
%%%%%%%%%%%%%%%%%%%%%%%%%%%%%%%%%%%%%%%
%$\lambda_{1}$ [eV\,\AA]&$\lambda_{2}$ [eV\,\AA]&$\lambda_{3}$ [eV\,\AA]\\\hline
% 0.012 &  0.009 & -0.015\\
% \hline
%%%%%%%%%%%%%%%%%%%%%%%%%%%%%%%%%%%%%%%
%\end{tabular}
%\end{table}
%%%%%%%%%%%%%%%%%%%%%%%%%%%%%%%%%%%%%%%

Finally, one can argue that the observed spin-orbit coupling in phosphorene does not originate from the proximity-induced interaction with the strong spin-orbit coupling material, WSe$_2$ ML, but is a consequence of the broken symmetry of the phosphorene monolayer. To test this assumption,  we compare the previously calculated spin-orbit coupling parameters with the case of the phosphorene ML, by removing the WSe$_2$ ML from the self-consistent calculation and keeping the coordinates of phosphorene ML obtained within the heterostructure relaxation, being the mechanism responsible for breaking the phosphorene's symmetry.
In this case, the fitting of the spin-orbit Hamiltonian~\eqref{Heff} to the DFT data gives us the following parameters: $\lambda_1^{\rm P}=-0.00065$ eV\,\AA, $\lambda_2^{\rm P}=0.0014$ eV\,\AA, and $\lambda_3^{\rm P}\approx 0$, confirming the dominant role of the proximity-induced spin-orbit coupling effect. Note that the obtained values obey a similar trend ($|\lambda_1^{\rm P}|<|\lambda_2^{\rm P}|$; $\lambda_3^{\rm P}=0$), and are of the same order of magnitude as 
Rashba spin-orbit parameters of phosphorene in strong electric fields ($\propto$\,V/nm)~\cite{KGF16}.

%%%%%%%%%%%%%%%%%%%%%%%%%%%%%%%%%%%%%%%%%%%%%%%%%%%%
\begin{table}[t]
\caption{Spin-orbit coupling parameters $\lambda_{\rm 1/2/3}$ obtained using different vdW corrections: Grimme-D2, Grimme-D3 correction with 
Becke and Johnson damping, Tkachenko-Scheffler, and the non-local rvv10. The calculated average distance $d$ between the top selenium and bottom phosphorene plane of the relaxed heterostructure is also given.}\label{vdWinfluence}
\centering
\small
\setlength{\tabcolsep}{8pt}
\renewcommand{\arraystretch}{1.2}
\begin{tabular}{|c|c|c|c|c|}
\hline
%%%%%%%%%%%%%%%%%%%%%%%%%%%%%%%%%%%%%%%
vdW & $\lambda_{1}$ [eV\,\AA]&$\lambda_{2}$ [eV\,\AA]&$\lambda_{3}$ [eV\,\AA]& d [\AA]\\\hline
D2 &0.012 &  0.009 & -0.015&3.31\\\hline
D3BJ &0.013 &  0.010 & -0.017&3.24\\\hline
rvv10 &0.008 &  0.009 & -0.011&3.41\\\hline
TS &0.005 &  0.009 & -0.006&3.66\\\hline
%%%%%%%%%%%%%%%%%%%%%%%%%%%%%%%%%%%%%%%
%%%%%%%%%%%%%%%%%%%%%%%%%%%%%%%%%%%%%%%
\end{tabular}
\end{table}
%%%%%%%%%%%%%%%%%%%%%%%%%%%%%%%%%%%%%%%

Since we have shown that the observed values of the spin-orbit coupling are obtained by transferring the strong spin-orbit coupling from WSe$_2$ to P via the proximity effect, we additionally analyze the influence of the different vdW correction on $\lambda$'s, which can affect the distance between the ML after relaxation and the interaction between the heterostructure constituents. Besides the Grimme-D2 vdW correction %used above, 
we additionally focus on the Grimme-D3 vdW correction~\cite{GAE+10} with 
Becke and Johnson damping (D3BJ)~\cite{GEG+11}, Tkachenko-Scheffler (TS)~\cite{TS09}, and the non-local rvv10~\cite{VV10,SGG13} vdW corrections. The results are gathered in Table~\ref{vdWinfluence}, where we additionally present the averaged distance between the top selenium and bottom phosphorene plane.% besides the spin-orbit parameters. 
When compared to the averaged distance 3.31\,\AA\, in the Grimme-D2 case, it can be concluded that D3BJ/rvv10 slightly decreases/increases the distance, while in the case of the TS vdW correction, the distance increase is more pronounced~($\approx 10\%$). Comparison of the obtained spin-orbit parameters shows that spin-dependent parameters obtained when using Grimme-D2 and D3BJ vdW corrections are in excellent agreement, while there is a slight/significant decrease of parameters $\lambda_1$ and $\lambda_3$ in the rvv10/TS case. Also, for all vdW corrections, the $\lambda_2$ parameter is basically unaffected.
Thus, we can conclude that the spin-orbit proximity effect is dependent on the choice of the vdW correction, but the qualitative picture and the order of magnitude of the effect remain unimpacted.

\subsection{Twist modification of proximity-induced spin-orbit coupling: an example of $60^{\rm o}$ twist angle}
Strong proximity-mediated transfer of a spin-orbit coupling from WSe$_2$ to phosphorene suggests that 
a relative change of WSe$_2$ band structure with respect to phosphorene by means of a twist could have a
significant impact on the spin texture in phosphorene.
We test this assumption by analyzing the P/WSe$_2$ heterostructure in which the WSe$_2$ monolayer is twisted for an angle of $60^{\rm o}$ with respect to phosphorene. The WSe$_2$ ML within the new heterostructure has the same number of atoms and is strained for the equal percentage as in Sec.~\ref{Heterostructure}; thus, it was possible to use the same parameters as before to perform the necessary DFT calculations. After fitting the model Hamiltonian~\eqref{Heff} to the DFT data, we obtain the following spin-orbit coupling parameters:  $\lambda_1=0.010$ eV\,\AA, $\lambda_2=0.010$ eV\,\AA, and $\lambda_3=0.015$ eV\, \AA. When compared to the values obtained in the zero twist-angle case, we can notice that a small change in parameters $\lambda_1$ and $\lambda_2$ is followed by the sign change of the $\lambda_3$ parameter. The sign change of the $\lambda_3$, corresponding to the $k_x \sigma_z$ spin-orbit coupling term, can be directly connected to the fact that, instead of the $\Gamma$K branch, the $\Gamma$K' branch of WSe$_2$ is located on the $\Gamma$X line of the phosphorene Brillouin zone. Since at the ${K}$ and ${K}'$ points, the corresponding energies are equal and connected via time-reversal symmetry $\Theta$, $\Theta E_{\ket{{K}+}}=E_{\ket{{K'}-}}$, where $\ket{\pm}$ corresponds to spin wavefunction with $s_z=\pm 1/2$ spin expectation value (we remind that spins in WSe$_2$ monolayer are locked in the out-of-plane direction),
hybridization of phosphorene bands with WSe$_2$ bands via the spin split branch with $s_z=\pm1/2$ spin expectation value will be transferred to the branch $s_z=\mp1/2$ with the opposite spin. The fact that the $k_x\sigma_z$ term is locked to the valley of WSe$_2$ MLs suggests that this term is related to the valley-Zeeman spin-orbit coupling induced by the proximity effect in the studied heterostructure.

\section{Conclusions}\label{Conclusions}
We analyzed the proximity-induced spin-orbit coupling effects in a heterostructure made of phosphorene and WSe$_2$ monolayer. Giant spin splitting of WSe$_2$ valence bands 
motivated us to focus on the hole spin physics in phosphorene where, due to the broken inversion symmetry, spin splitting of the bands can occur. 
We discovered a significant proximity-induced spin-orbit coupling in the top valence band of phosphorene, whose origin is attributed to the strong hybridization with the WSe$_2$ spin split bands close to the $\Gamma$ point.
An effective spin-orbit coupling Hamiltonian model compatible with the ${\bf C}_{1{\rm v}}$ symmetry of the heterostructure is derived, and the spin-orbit parameters that fit the obtained data from ab-initio calculations to the model Hamiltonian are determined.  
By comparing the obtained parameters with the spin-orbit coupling values with group-IV monochalcogenide monolayers, representing the ferroelectrics with phosphorene-like atomic structure, we concluded that phosphorene is transformed into weak spin-orbit coupling material. Still, compared to electric field-induced Rashba spin-orbit coupling, the proximity-induced spin-orbit coupling is an order of magnitude larger.
Finally, we showed that the twist angle can influence the spin-orbit proximity effect in a studied material. More precisely, for the twist angle of 60$^{\rm o}$, we reported a sign change of the out-of-plane spin-orbit field, followed by a sizable modification of the in-plane spin-orbit texture. 
Although the discovered spin splitting due to proximity-induced spin-orbit coupling is much less than the room temperature, it should not represent a limitation for spintronics application, as demonstrated recently in a bilayer graphene/WSe$_2$ heterostructure~\cite{AHF+21}, having the meV size of spin splitting of bilayer graphene bands also.
Thus, the presented study shows that structures with incompatible symmetries can be used to generate spin textures different from the more commonly studied composites made of graphene and transition metal dichalcogenides, opening a playground for novel materials that can be used either as a target material or as a substrate in van der Waals heterostructures important for spintronics application.
%%%%%%%%%%%%%%%%%%%%%%%%%%%%%%%%%%%%%%%%%%%%%%%%%%%%%%%%%%%%%%%%%%%%%%%%%%%%%%%%%%%%%%%%%%%%
\acknowledgments
M.M. acknowledges the financial support
provided by the Ministry of Education, Science, and Technological Development of the
Republic of Serbia and DAAD Research Grant 57552336. This project has received
funding from the European Union's Horizon 2020 Research and Innovation Programme under the Programme SASPRO 2 COFUND Marie Sklodowska-Curie grant agreement No. 945478.
%%%%%%%%%%%%%
M.G.~acknowledges financial support provided by Slovak Research and Development Agency provided under Contract No. APVV-SK-CZ-RD-21-0114 and by the Ministry of Education, Science, Research and Sport of the Slovak Republic provided under Grant No. VEGA 1/0105/20 and Slovak Academy of Sciences project IMPULZ IM-2021-42 and project FLAG ERA JTC 2021 2DSOTECH.
%%%%%%%%%%%%%
M.K.~acknowledges financial support provided by the National Center for Research and Development (NCBR) under the V4-Japan project BGapEng V4-JAPAN/2/46/BGapEng/2022. 
%%%%%%%%%%%%%
I.{\v S}~acknowledges financial support by  APVV-21-0272, VEGA 2/0070/21, VEGA 2/0131/23, and by H2020 TREX GA No. 952165 project.
%%%%%%%%%%%%%%%%%%%%%%%%%%%%%%%%%%%%%%%%%%%%%%%%%%%%%%
J.F. acknowledges support from Deutsche Forschungsgemeinschaft (DFG, German Research Foundation) SFB 1277 (Project-ID 314695032, project B07), SPP 2244 (Project No. 443416183), and of the European Union Horizon 2020 Research and Innovation Program under Contract No. 881603 (Graphene Flagship) and FLAG-ERA project 2DSOTECH.
The authors gratefully acknowledge the Gauss Centre for Supercomputing e.V. (www.gauss-centre.eu) for funding this project by providing computing time on the GCS Supercomputer SuperMUC-NG at Leibniz Supercomputing Centre (www.lrz.de).

%%%%%%%%%%%%%%%%%%%%%%%%%%%%%%%%%%%%%%%

\end{document}